\documentstyle[12pt]{article}
\begin{document}
\thispagestyle{empty}
\begin{center}
{\LARGE \tt \bf Photon mass and Cosmological constant bounds from spacetime torsion}
\end{center}
\vspace{5cm}
\begin{center}
{\large by L.C. Garcia de Andrade\footnote{Departamento de F\'{\i}sica 
Te\'{o}rica -Instituto de F\'{\i}sica- UERJ - CEP: 22.022 Rua S\~{a}o 
Francisco Xavier, 524 - Maracan\~{a} RJ - Rio, Brasil.}}
\end{center}
\begin{abstract}
{\large Photon mass and Cartan contortion bounds recently obtained from tiny Lorentz violation  observations in cosmology are used to find a limit of ${\lambda}\le 10^{-4}{\alpha}$ for the massive photon-torsion dimensionless coupling. Here ${\alpha}$ represents the fine-structure constant. A gauge invariant Proca electrodynamics in spacetime endowed with torsion in de Sitter spacetime is used to obtain an upper bound for the present value of the cosmological constant given by ${\Lambda}\le 10^{-56} cm^{-2}$. This result is obtained in regions of the universe where the photons are massless. A relation between the contortion, photon mass and the radius of the universe is obtained. The Proca electrodynamics with torsion and the radius of the universe allow us to place more stringent bounds for the photon mass of $m_{\gamma}{\le} 10^{-44} GeV$ which is only two orders of magnitude lower than the astronomical bounds given by the PARTICLE DATA GROUP (PDG). We also show that charge is locally conserved in de Sitter spacetime with torsion and that plane waves are shown to be damping by contortion inhomogeneities while dispersion is isotropic and therefore Proca-Cartan photons do not violate Lorentz invariance.}
\end{abstract}
\newpage
\section{Introduction}
\hspace{0.6cm}
Recent results by Prokopec et al \cite{1} show that Proca electrodynamics for the photon mass can be obtained from Hartre-Fock approximation to scalar QED during inflation, which allows us to place upper bounds for the photon mass. Also recently Luo et al \cite{2} photon mass bounds from rotatable torsion balance. Also recently gauge invariant vectorial photon mass in axionic electrodynamics, where the use of torsion is not mandatory \cite{3} was obtained by M. Haugan and C. L\"{a}mmerzahl \cite{4}. Of course their electrodynamics is a test field theory for the Maxwell field and does not represent the dynamics of the Proca field. In this paper we follow reference \cite{1} extending it to spacetime with torsion. In this paper we follow analogous computation to Andrianov et al \cite{5} where they use a spinor QED modification to include Lorentz and CPT breaking axial-vector. Left and right polarized Proca-Cartan photons are obtained. Recently E. Aldelberger et al \cite{6} showed that by considering a photon mass arising from a Higgs effect (longitudinal Goldstone boson) it is possible to destroy photon mass from vortices. They also show that photon mass vanish inside galaxies while become massive outside it. Such a phenomena of disapearence of photon mass can be thought here in the universe scale where torsion and inflation are shown to provide a mechanism to the photon mass decay. Another motivation for our work stems from the fact that Andrianov et al \cite{5} considered the possibility that a small modification of spinor QED allows for the presence of massive photons where Lorentz and CPT non-covariant kinetic terms are present. The paper is organized as follows: In section $2$ we present the Proca-Cartan electrodynamics and place bounds on the massive photon-torsion coupling constant. In section $3$ we compute the conditions Cartan contortion and the cosmological constant should satisfy in order that the universe be free of massive photons. In this section we also place upper bounds for the cosmological constant in the present universe. In section $4$ the dispersion relation of plane waves is presented. In this section we also compute the photon mass bounds from known values for torsion obtained from Lorentz violation in cosmology \cite{7} and also the torsion bounds from astrophysical stringent limits on the photon mass. Section $5$ deals with the conclusions and discussions.
\section {Proca electrodynamics in torsioned spacetime}
Earlier Garcia de Andrade and Sivaram \cite{8} has proposed a non-minimal coupling Einstein-Cartan-Proca Lagrangian 
\begin{equation}
L= \sqrt{-g}[(1+{\lambda}A^{2})R(\Gamma)-\frac{1}{4}F_{ij}F^{ij} + J^{i}A_{i}]
+ KL_{m}
\label{1}  
\end{equation}
Besides the torsion vector may be interpreted as a massive (Proca) field and the Ricci-Cartan scalar is represented by $R({\Gamma})$, where ${\Gamma}$ is the non-Riemannian connection. Here $A^{2}:= A_{i}A^{i}$ where $(i,j=0,1,2,3)$ and $J^{i}$ is the electromgnetic vector current. The constant ${\lambda}$ represents the massive photon-torsion coupling which we shall determine in this Letter. $L_{m}$ denotes the matter Lagrangean. Here we assume that the electromagnetic field tensor $F_{ij} = {\partial}_{i}A_{j} - {\partial}_{j} A_{i}$ is semi-minimally coupled to the gravitational field. This was first proposed by de Sabbata and Gasperini \cite{9} in the context of QED virtual photons associated with fermion anti-fermion pairs. In their approach the Maxwell tensor  $F^{\mu \nu}$ does not couple with torsion through minimal coupling but couples with torsion only through the term ${\lambda}R({\Gamma})A^{2}$ on the non-minimally coupled Lagrangean (\ref{1}). The photon mass comes from a dynamical relation between torsion and electromagnetic vector potential indicating that at least in first approximation the photon mass $m_{\gamma}$ is connected to Ricci-Cartan scalar. The electromagnetic equations are obtained by variation Lagrangean (\ref{1}) w.r.t. $A_{i}$. Thus
\begin{equation}
{\nabla}_{j}F^{ij}= J^{i} -{\lambda}R({\Gamma})A^{i}
\label{2}
\end{equation}
where ${\nabla}_{j}$ is the Riemannian covariant derivative operator. To obtain the Proca equation we simply use the definition of the electromagnetic field with non-minimal coupling to torsion 
\begin{equation}
{\Box}A^{i}+{\lambda}R({\Gamma})A^{i}=0
\label{3}
\end{equation}
The Riemannian D'Lambertian is ${\Box} = {\nabla}_{i}{\nabla}^{i}$. Before we proceed let us take sometime to an important question, namely the of charge conservation in spacetimes with torsion \cite{10}. In our case this is simple if we redefine the currents
\begin{equation}
{J^{*}}^{i} = J^{i}-{\lambda}R({\Gamma}) A^{i}
\label{4}
\end{equation}
therefore charge would be conserved now if ${{{\nabla}^{*}}_{i}}{J^{*}}^{i}=0$. Here ${{{\nabla}^{*}}_{i}}$ denotes the Riemann-Cartan covariant derivative operator. Note that if we also assume that ${{\nabla}_{i}}{J^{i}}=0$ we are constraint to
\begin{equation}
A^{i}{\partial}_{i}(R({\Gamma}))=0
\label{5}
\end{equation}
where we use the Riemannian Lorentz gauge condition ${\nabla}_{i}A^{i}=0$. But even in this case this is enough for our purposes in this Letter since for the de Sitter spacetime in first approximation of torsion we have
\begin{equation}
A^{i}{\partial}_{i}[R({\left\{\right\}})+{\partial}_{0}K^{0}] = 0
\label{6}
\end{equation}
because in de Sitter universe the Riemann-Ricci tensor $R({\left\{\right\}})=12{\Lambda}$ and this left us with the expression for contortion that vanishes since we may use the gauge $(A^{0}=0,\vec{A})$. Therefore as far as our purposes are concerned the presence of photon mass in spacetime with torsion does not violate local conservation of charge. From the Proca-Cartan equation one may write 
\begin{equation}
{m_{\gamma}}^{2}= {\lambda}[12{\Lambda}+{\partial}_{0}K^{0}]
\label{7}
\end{equation}
where $K^{0}$ is the time-component of the contortion vector $K^{i}= {\epsilon}^{ijkl}K_{jkl}$ wher $K_{ijk}$ is the contortion tensor. Since ${\lambda}$ is a constant we shall assume that the cosmological constant here vanishes to compute it from well-known data of photon mass bounds \cite{11}. Making use of the bound $m_{\gamma}\le 10^{-42} GeV$ one obtains the following inequality
\begin{equation}
{\lambda}{\partial}_{0}K^{0}\le 10^{-42} GeV
\label{8}
\end{equation}
and
\begin{equation}
{\lambda}K^{0}\le 10^{-42}R_{U} GeV
\label{9}
\end{equation}
where $R_{U}$ represents the Hubble radius of the universe given by $R_{U}=10^{28}cm$. Substitution of this value into expresion (\ref{9}) yields ${\lambda}K^{0}\le 10^{-38}eV$. From this expression and the upper bound of contortion $K^{0}\le 10^{-32} eV$ obtained by L\"{a}mmerzahl we obtain the following result ${\lambda}\le 10^{-6}$ for the massive photon-torsion coupling. To better comparison with QED one may express this result in  terms of the fine-structure constant ${\alpha}$ as ${\lambda}\le 10^{-4}{\alpha}$. Note that this coupling constant is low w.r.t. to the Maxwell theory. This result would be expected since the coupling torsion is in general a very weak field compared to other fields like the electromagnetic field and Einstein curvature gravitational field. One notes that this limit is however not so stringent as ours. In their lagrangean the Ricci tensor in formula (\ref{1}) is absent.
\section{Cosmological constant from torsion}
Since as we mention before the massive photons could be destroyed by vortices and vortices can be easily associated with torsion, would be natural to think that in same way torsion and inflation could be also responsible by the absence of massive photons in expanding universes. Therefore to check for the physical effects of this assumption we assume from the beginning that we can put the photon mass equal to zero in equation (\ref{7}). An immediate result is 
\begin{equation}
{\Lambda}=-\frac{{\partial}_{0}K^{0}}{12}
\label{10}
\end{equation}
Since torsion is a positive function we conclude that if ${\Lambda}<0$ contortion does not decrease in time in this universe. However, this is an unphysical assumption since otherwise we would have already detected torsion in the present universe. Therefore our assumption of a positive cosmological constant is consistent with a universe with torsion \cite{9} devoided of massive photons. Let us now compute the cosmological constant from the ${\lambda}$ value obtained in the previous section and the expression
\begin{equation}
{\Lambda}=-\frac{K^{0}}{12}{R_{U}}^{-1}
\label{11}
\end{equation}
which is derived from (\ref{10}). For the value of $K^{0}$ we use again the bound computed by Garcia de Andrade \cite{7} given by $K^{0}\le 10^{-32} eV$. A quick computation yields ${\Lambda}\le 10^{-56} cm^{-2}$ for the present value of the cosmological constant. Of course this result gives yet more support to cosmologists which do not take the cosmological constant into account in the present cosmological models. 
\section{Damping of plane waves and photon mass bounds from torsion}
Recently Haugan and L\"{a}mmerzahl \cite{4} have proposed a test theory for Maxwell field based on a vectorial photon mass. Their Maxwell generalised equation is gauge invariant and the eletromagnetic plane wave present anisotropy dispersion and damping. In this section we show that only isotropic dispersion appears in our case while contortion is a source of damping for the electromagnetic plane wave. Let us consider the Proca equation and the electromagnetic potential given by
\begin{equation}
A^{r}={A_{0}}^{r} e^{i(k_{l}x^{l})}
\label{12}
\end{equation}
where the  amplitude ${A_{0}}^{r}$ is not necessarily constant throughout spacetime. Substitution of expression (\ref{12}) into Proca equation yields
\begin{equation}
{\Box}{A_{0}}^{r}+i[k_{l}k^{l}{A_{0}}^{r}+2 k^{l}{\partial}_{l}{A_{0}}^{r}]+{\lambda}{\partial}_{0}K^{0} {A_{0}}^{r}=0
\label{13}
\end{equation}
To simplify matters we consider now that the amplitude of electromagnetic waves is constant, namely plane electromagnetic waves, and that the Riemannian Ricci tensor also vanishes. This Riemann-flat case is justified since in this section we are interested more on the torsion rather than curvature effects. This reduces equation (\ref{13}) to
\begin{equation}
ik_{l}k^{l}+{\lambda}{\partial}_{0}K^{0}=0
\label{14}
\end{equation}
Solving the dispersion relation (where $k^{l}k_{l}={\omega}^{2}-{\vec{k}}^{2}$) one obtains
\begin{equation}
{\omega}=\pm{|\vec{k}|}\sqrt{1+i\frac{{\lambda}{\partial}_{0}K^{0}}{|\vec{k}|^{2}}}
\label{15}
\end{equation}
\begin{equation}
{\omega}\cong\pm{|\vec{k}|}[{1+i\frac{{\lambda}{\partial}_{0}K^{0}}{2|\vec{k}|^{2}}}]
\label{16}
\end{equation}
From this expression we note that contortion induces a damping into the electromagnetic plane wave but contrary to previous test theories of the Maxwell field there is no anisotropy dispersion of plane waves on the torsioned background. Therefore here we do not have Lorentz violation. For other test Maxwell theories which possess Lorentz violation the we refer to reference \cite{12}. Now let us  assume that the cosmological constant vanishes in the actual universe and that locally only torsion and massive photons matters in this pedestrian cosmological model. Thus from expression (\ref{7}) one obtains
\begin{equation}
{m_{\gamma}}^{2}= {\lambda}K^{0}{R_{U}}^{-1}
\label{17}
\end{equation}
Substitution again of the bounds for contortion one obtains the following upper bound for photon mass $m_{\gamma}\le 10^{-44}GeV$. This bound is two orders of magnitude more stringent than the best value known so far which is $m_{\gamma}\le 10^{-42} GeV$. The PDG presented $m_{\gamma}\le 10^{-25} GeV$ in his 2002 data table. 
\section{Discussion and conclusions}
In this brief report we have shown that a Proca electrodynamics in de Sitter universe with torsion may provide a simple model and laboratory to compute upper bounds for some of the most intriguing physical objects in nature, namely the cosmological constant and massive photons ,not to mention Cartan geometrical torsion itself. On the other hand to base this model on a cosmological model with torsion allows us to conclude that a stronger basis for this kind of cosmology can be  given as a physical theory. Besides these data other physical support to cosmology with torsion have been established recently through the investigation of Einstein-Cartan cosmologies with and without propagating torsion by Lasenby et al \cite{13} and myself \cite{14} using COBE and CHANDRA data. Our model may also serve to a departuring point to investigate more complicate models like Friedmann or anisotropic models in the Early universe where torsion, cosmological constant and even massive photons may play a more fundamental role. Besides models like the Einstein-Cartan-Kalb-Rammond developed by SenGupta et al \cite{15} could be generalized to allow for photon mass spectrum. The extremely low value for torsion used in this paper in the present universe was alternatively recentely explained by Mukhopadhyaya et al \cite{16} on the basis of Randall-Sundrum scenario. Bounds on torsion parameters have been also recently given by Mahanta and Raychaudhuri \cite{17}. Photon mass models with torsion in the context of Kaluza-Klein theory has been recently considered by C. Kohler \cite{18}. In his approach a five-dimensional unification between the cosmological constant and photon mass in terms of Einstein-Cartan theory is also given, however, his relation between photon mass and cosmological constant does not depend on contortion and no attempt is made of placing upper bounds neither to photon mass nor to cosmological constant or even torsion.
\section*{Acknowledgements}
I am very much indebt to Professors F. W. Hehl, S.P. Sorella, R. Ramos and Dr. C. L\"{a}mmerzahl for useful discussions on the subject of this paper. One of (GA) would  like to thank CNPq. (Brazilian Research Council) and Universidade do Rio de Janeiro for financial support.

\end{document}